\documentclass[prc,twocolumn,preprintnumbers,superscriptaddress]{revtex4-1}

\usepackage{graphicx}
\usepackage{bm}
\usepackage{amsmath}
\usepackage{dcolumn}

\def\bea {\begin{eqnarray}}
\def\eea {\end{eqnarray}}
\def\be {\begin{equation}}
\def\ee {\end{equation}}
\def\ben{\begin{enumerate}}
\def\een{\end{enumerate}}
\def\bi{\begin{itemize}}
\def\ei{\end{itemize}}

\def\hyphen{{\mbox{-}}}

\def\2p1h{2p\hyphen 1h }
\def\3p2h{3p\hyphen 2h }

\begin{document}




\title{The half-life of $^{221}$Fr in Si and Au at 4K and at mK temperatures.}


\author{F. Wauters}
\affiliation{Instituut voor Kern- en Stralingsfysica, Katholieke Universiteit Leuven,
B-3001 Leuven, Belgium}
\author{B. Verstichel}
\altaffiliation[Present address: ]{Center for Molecular Modeling, Ghent University, Technologiepark 903, 9052
Zwijnaarde, Belgium}
\affiliation{CERN, CH-1211 Gen\`{e}ve 23, Switzerland}
\affiliation{Instituut voor Kern- en Stralingsfysica, Katholieke Universiteit Leuven,
B-3001 Leuven, Belgium}
\author{M. Breitenfeldt}
\affiliation{Instituut voor Kern- en Stralingsfysica, Katholieke Universiteit Leuven,
B-3001 Leuven, Belgium}
\author{V. De Leebeeck}
\affiliation{Instituut voor Kern- en Stralingsfysica, Katholieke Universiteit Leuven,
B-3001 Leuven, Belgium}
\author{V. Yu. Kozlov}
\affiliation{Instituut voor Kern- en Stralingsfysica, Katholieke Universiteit Leuven,
B-3001 Leuven, Belgium}
\author{I. Kraev}
\affiliation{Instituut voor Kern- en Stralingsfysica, Katholieke Universiteit Leuven,
B-3001 Leuven, Belgium}
\author{S. Roccia}
\affiliation{Instituut voor Kern- en Stralingsfysica, Katholieke Universiteit Leuven,
B-3001 Leuven, Belgium}
\author{G. Soti}
\affiliation{Instituut voor Kern- en Stralingsfysica, Katholieke Universiteit Leuven,
B-3001 Leuven, Belgium}
\author{M. Tandecki}
\affiliation{Instituut voor Kern- en Stralingsfysica, Katholieke Universiteit Leuven,
B-3001 Leuven, Belgium}
\author{E. Traykov}
\affiliation{Instituut voor Kern- en Stralingsfysica, Katholieke Universiteit Leuven,
B-3001 Leuven, Belgium}
\author{S. Van Gorp}
\affiliation{Instituut voor Kern- en Stralingsfysica, Katholieke Universiteit Leuven,
B-3001 Leuven, Belgium}
\author{D. Z\'{a}kouck\'{y}}
\affiliation{Nuclear Physics Institute, ASCR, 250 68 \v{R}e\v{z}, Czech Republic}
\author{N. Severijns}
\affiliation{Instituut voor Kern- en Stralingsfysica, Katholieke Universiteit Leuven,
B-3001 Leuven, Belgium}
\email{nathal.severijns@fys.kuleuven.ac.be}

\date{\today}


\begin{abstract}

The half-life of the $\alpha$~decaying nucleus $^{221}$Fr was determined in different environments, i.e. embedded in Si at 4~K, and embedded in Au at 4~K and about 20~mK. No differences in half-life for these different conditions were observed within 0.1~\%. Furthermore, we quote a new value for the absolute half-life of $^{221}$Fr of t$_{1/2}$ = 286.1(10)~s, which is of comparable precision to the most precise value available in literature.

\end{abstract}

%
%



\maketitle

\section{Introduction}

The half-life of $\alpha$~and $\beta$ decaying nuclei have always been considered as constant, independent of their surroundings. Recently, however, it has been claimed that the half-life of a radioactive isotope would change if embedded in a metallic host. Using the Debye plasma model applied to quasi-free metallic electrons \cite{Rolfs2006,Raiola2005,Kettner2006}, a $1/\sqrt{T}$ dependence for the screening energy U$_D$ was predicted. When cooled down to a few Kelvin, the half-life of $\beta^-$($\beta^+$) decaying nuclei would then increase (decrease) by several tens of percent and the half-life of $\alpha$~decaying nuclei would even be shortened by several orders of magnitude. This hypothetical mechanism was proposed as a possible solution for long lived transuranic waste produced by fission reactors \cite{Rolfs2006}.
\\
A series of subsequent experiments claimed to have observed changes in the half-lives of both $\beta^{\mp}$ and $\alpha$~ decays, although the effects were less profound than predicted by the Debye plasma model. The half-life of $\alpha$-decaying $^{210}$Po implanted in copper was reported to shorten by 6.3(14)~\% when cooled down to 12~K \cite{Raiola2007}, the half-life of $\beta^+$ decaying $^{22}$Na embedded in Pd was observed to be 1.2(2)~\% shorter at 12~K \cite{Limata2006}, the half-life of $\beta^-$ decaying $^{198}$Au embedded in Au was observed to be 4(2)~\% longer at 12~K \cite{Spillane2007}, and for $^{7}$Be (EC-decay), an increase of about 1~\% was reported \cite{Wang2006}.
Note that electron capture decay rates may depend on the material hosting the radioactive isotope via small modifications of the electron density around the EC decaying nucleus (see e.g. \cite{Nir-El2008} and references therein).
In contrast to the above results, several experiments carried out at a later stage on the $\beta$ decaying isotopes $^{198}$Au \cite{Goodwin2009,Kumar2008,Ruprecht2008,Ruprecht2008b}, $^{22}$Na \cite{Ruprecht2008,Ruprecht2008b}, $^{64}$Cu \cite{Fallin2008}, and $^{74}$As \cite{Farkas2009}, and on the EC decaying $^7$Be \cite{Kumar2008}, did not observe any changes of the half-lives up to the permille level when these isotopes were embedded in a metallic environment and cooled down to 10-20~K.
\\
As to $\alpha$~decay, it was demonstrated theoretically that taking into account not only the electron screening effect on the $\alpha$~particle's tunneling potential, but also on its binding energy inside the nucleus, the half-life would not significantly change \cite{Zinner2007}. Further, it was argued \cite{Severijns2007,Stone2007b} that if the Debye plasma model would be applicable to $\alpha$~decays, the effect should have been observed previously already in Low Temperature Nuclear Orientation experiments (LTNO) \cite{Stone1986}.
In this type of experiments radioactive isotopes are implanted in a metallic foil, typically Fe, and cooled to milliKelvin temperatures. It was claimed \cite{Spillane2007,Raiola2007} that no appreciable effect on nuclear half-lives can be observed in LTNO experiments, as the radioactive ions implanted at typically 60 keV would end up in the oxidized surface layer of the sample foil, which acts as an insulator. This claim is clearly incorrect as the LTNO technique is precisely based on the fact that the radioactive nuclei end up at substitutional sites in a pure bcc Fe lattice, where they experience a unique hyperfine interaction. Nuclei that end up in an oxide surface layer cannot be oriented and thus also do not show anisotropic emission. To avoid a surface oxide layer, the foils used in LTNO measurements are carefully prepared by polishing and annealing procedures prior to inserting them in the vacuum of the setup \cite{vanWalle1986,Herzog1985}. Substitutional fractions of about 70~\% to 95~\% are generally observed for 60 keV implantations into a cold (i.e. 4 K or lower) Fe foil (e.g. \cite{Stone2007,Severijns2005,Schuurmans1999,Wouters1990}). In LTNO experiments on $\alpha$~decaying isotopes implanted in Fe, no half-life changes at the percent level were observed between room temperature and 1~K \cite{Stone2007b}, and between 4~K and 50~mK \cite{Severijns2007}.
\\
Here, we report a dedicated half-life measurement of the $\alpha$~decaying nucleus $^{221}$Fr implanted in Si at 4~K, and in Au at 4~K and at 20~mK. The three different half-life values obtained agree witch each other within 0.1~\% and also agree with the literature room temperature value.
Previously, it was shown that the half-life of $^{221}$Fr is constant to less than a percent at room temperature, regardless to its chemical environment \cite{Jeppesen2007}.

\section{The experiment}

The $^{221}$Fr activity was produced at the ISOLDE/CERN facility by spallation reactions induced by a 1.4~GeV proton beam impinging on a UC$_x$ target \cite{Evensen1997}. After diffusing out of the target, the $^{221}$Fr nuclei were ionized in a W surface ion source \cite{Wenander2003}, accelerated to 60~keV, and mass selected by the General Purpose Separator (GPS) \cite{Kugler2000}.
The radioactive ions were subsequently transported to the NICOLE LTNO setup \cite{Schosser1988,Wouters1990} where they were implanted into the cooled Au or Si sample. A SRIM calculation \cite{Ziegler2004} showed that the implantation depth of $^{221}$Fr ions with an energy of 60~keV is 80~${\text{\AA}}$ for Au and 330~${\text{\AA}}$ for Si, respectively. In previous LTNO experiments, it was shown that under these implantation conditions about 80~\% of the $^{221}$Fr ions end up in a fully metallic environment in a well-prepared Fe sample foil \cite{Schuurmans1999,Wouters1990b}.
\\
The decay $\alpha$~particles were detected with three 500~$\mu$m thick Si PIN diode detectors which were positioned at an angle of 15$^\circ$ relative to the surface of the sample foil and at an angle of 90$^\circ$ relative to each other. These type of detectors have previously been tested under the conditions of this experiment and showed good behavior \cite{Wauters2009}. A high efficiency coaxial germanium detector, place outside the cryostat, was used for $\gamma$~ray detection. Further, standard NIM electronics and a PC-based data acquisition system were used.
\\
The measurements consisted of a succession of implantation and data-taking periods, where in the latter every 30~s a spectrum was recorded during at least 10 $^{221}$Fr half-lives. First, nine such measurements were performed with an Au foil at milliKelvin temperatures, after which the temperature of the foil was raised to 4~K and another five measurements were carried out. Thereafter, the Au sample foil was removed and a Si sample was loaded into the system to carry out five measuring cycles at 4~K.
During all measurements, a 1~kHz pulser was used to monitor the dead time. Only data for which the dead time was below 5~\% and the shorter-lived beam contaminant $^{221}$Ra (t$_{1/2}$ = 28(2)~s \cite{AKOVALI1990}) had decayed, were considered for analysis.
\\
\begin{figure}[!ht]
\centering
\includegraphics[width = 0.95\columnwidth]{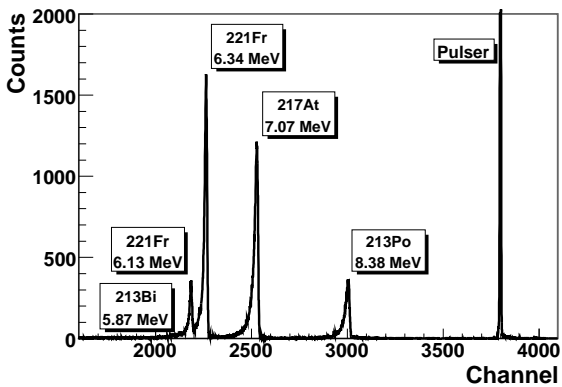}
\caption {A typical $\alpha$~spectrum, taken a few minutes after the end of the implantation of the mass 221 beam. The integral of the pulser peak was used to correct for dead time.} \label{fig:221Fr_alpha}
\end{figure}
A typical $\alpha$~spectrum is shown in Fig. \ref{fig:221Fr_alpha}. After the $^{221}$Ra activity has decayed, the only $\alpha$~lines appearing in the spectrum originate from the decay of $^{221}$Fr and its daughter nuclei. The $\alpha$~line at 6.34~MeV, which will be used to determine the half-life of $^{221}$Fr, showed an energy resolution of 32~keV (FWHM). The $\alpha$~lines of $^{217}$At and $^{213}$Po were broadened to respectively 43~keV and 57~keV because of the relocation of the recoiling daughter nuclei following the $\alpha$~decay of $^{221}$Fr and $^{217}$At, respectively, leading to a larger scattering for the decay $\alpha$ particles and even allowing part of nuclei to leave the host foil. Therefore, only the $\alpha$~lines of $^{221}$Fr, which was directly implanted, were considered for analysis.
\\
\begin{figure}[!ht]
\centering
\includegraphics[width = 0.95\columnwidth]{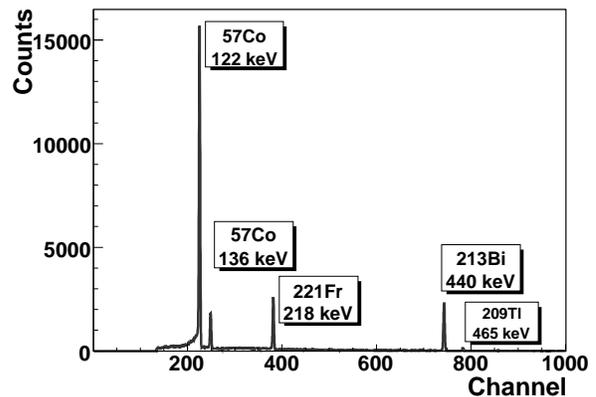}
\caption {Typical $\gamma$-ray spectrum. Apart form the two $\gamma$ lines of $^{57}$Co, only $\gamma$ lines originating from the decay of $^{221}$Fr or its daughter nuclei were observed. } \label{fig:221Fr_gamma}
\end{figure}
The $\gamma$ spectrum shown in Fig.~\ref{fig:221Fr_gamma} confirmed that, after the isotope $^{221}$Ra had decayed, $^{221}$Fr and its daughters were the sole activities present in the sample foil. The two $\gamma$ lines at 122~keV and 136~keV originate from the $^{57}$Co\underline{Fe} nuclear thermometer \cite{Marshak1986} which was soldered on to the back side of the sample holder so as to monitor the temperature in the milliKelvin range.

\section{Analysis and Results}

\begin{figure}[!htb]
\centering
\includegraphics[width = 0.8\columnwidth]{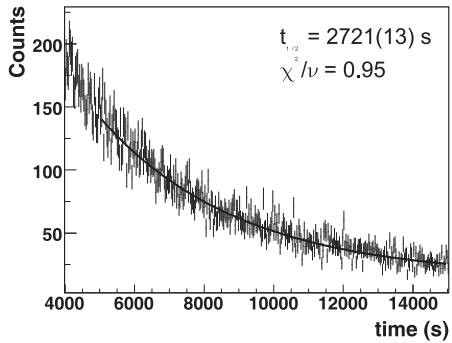}
\caption {A fit with a single exponential of the background of the $^{221}$Fr $\alpha$ peak integral. The resulting half-life coincides with the literature value of the half-life of $^{213}$Bi.} \label{fig:Bibgfit}
\end{figure}

\begin{figure}[!htb]
\centering
\includegraphics[width = 0.8\columnwidth]{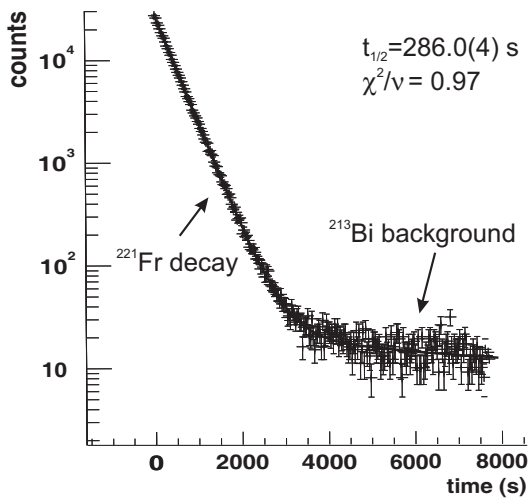}
\caption {Fit to the $^{221}$Fr half-life, t$_{1/2}$, of the decay curve of the combined $^{221}$Fr $\alpha$~peaks. The background decays according to the half-life of $^{213}$Bi (see Fig.~\ref{fig:Bibgfit}).} \label{fig:221Fr_decayFit_corr}
\end{figure}

\begin{figure}[!hbt]
\centering
\includegraphics[width = 0.9\columnwidth]{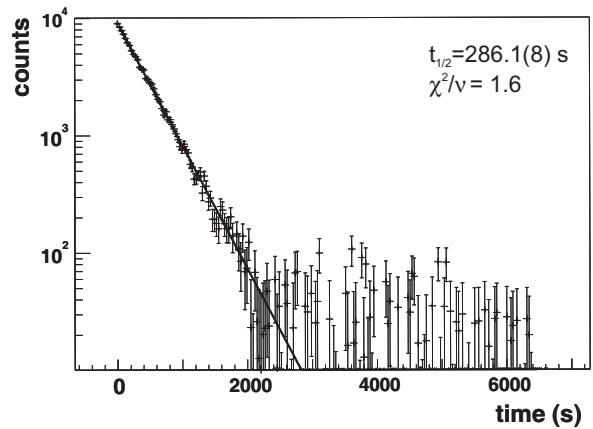}
\caption {Single exponential fit to the decay curve of the $^{221}$Fr 218~keV $\gamma$ line. } \label{fig:221Fr_GammadecayFit_corr}
\end{figure}

\begin{figure}[!htb]
\centering
\includegraphics[width = 0.8\columnwidth]{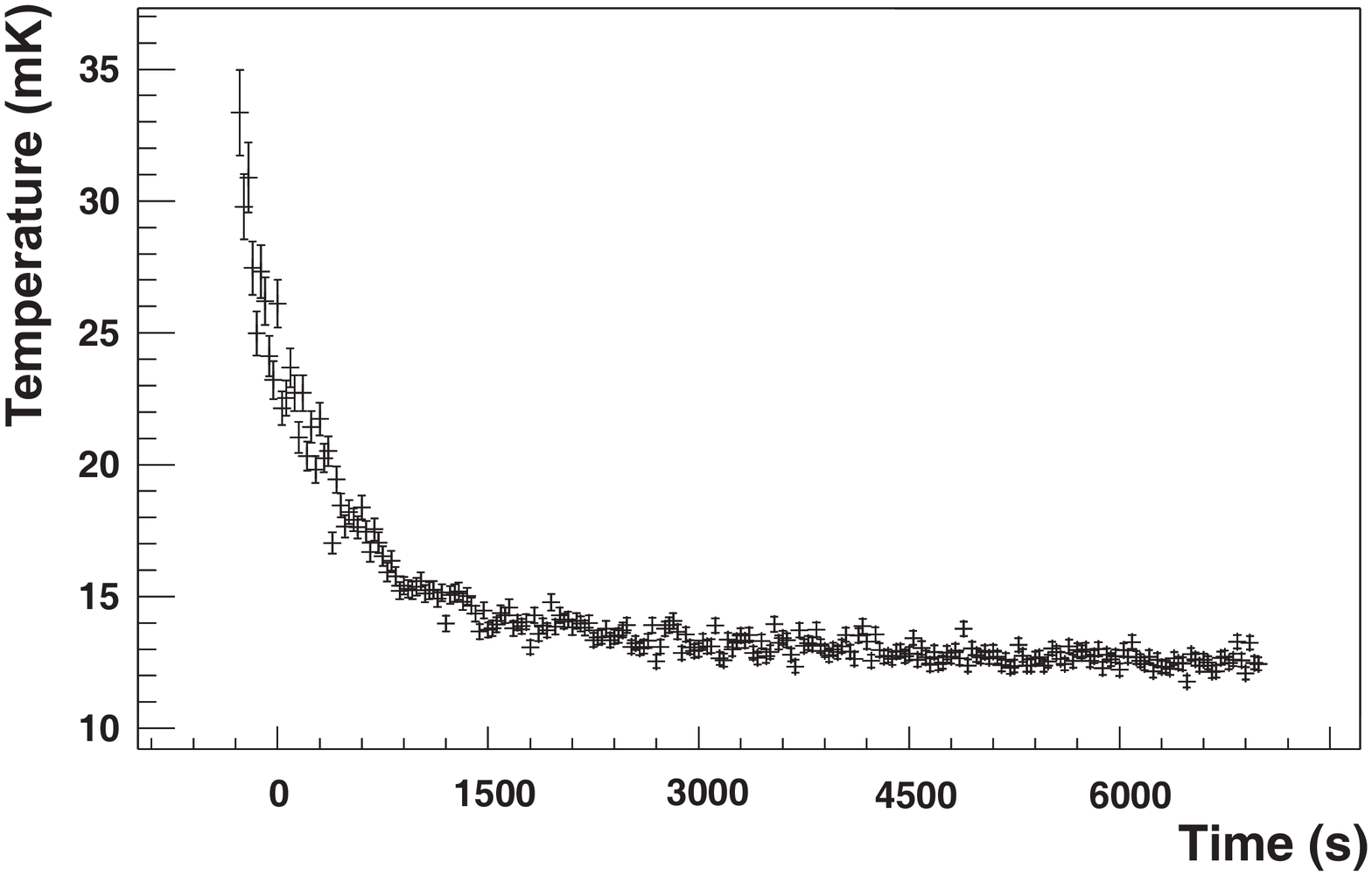}
\caption {Temperature evolution of the Au sample during one of the measurements at mK temperatures deduced from the $^{57}$Co\underline{Fe} nuclear orientation thermometer. During the implantation period, the sample warms to several tens of milliKelvin due to the power deposited by the mass 221 beam and the subsequent radioactive decay of the source that is gradually increasing in strength. About 200~s after the end of the implantation period, all of the short lived $^{211}$Ra activity is gone and the sample foil quickly cools down to below 20~mK, after which the half-life measurement was started.} \label{fig:Temperature_evolution}
\end{figure}

In order not to be sensitive to possible small gain shifts during the measurements, the $^{221}$Fr 6.34~MeV $\alpha$~peak was integrated using reasonably wide markers, including in fact both  $\alpha$~lines of $^{221}$Fr (at 6.34~MeV and 6.13~MeV \cite{AKOVALI2003}), as well as the weak 5.87~MeV $\alpha$~line of $^{213}$Bi \cite{NNDC1991Martin}. The background contribution to this integral originates from this $^{213}$Bi $\alpha$~line and from the long tail of the 8.38~MeV $\alpha$~line from the decay of $^{213}$Po \cite{NNDC1991Martin}, which has the same effective half-life as $^{213}$Bi. After correcting the integrals for the dead-time by normalizing to the pulser peak (see Fig.~\ref{fig:221Fr_alpha}), it was checked that the background indeed decays according to the literature value of the half-life of $^{213}$Bi, i.e. 2735(4)~s \cite{NNDC1991Martin} (Fig.~\ref{fig:Bibgfit}).
\\
\\
To fit the $^{221}$Fr decay curve, the following function was used:
\begin{equation}
Ae^{-\frac{(t+B)ln2}{C}}+D\left[ e^{-\frac{(t+B)ln2}{E}} - e^{-\frac{(t+B)ln2}{C}}\right]\;\;,
\label{eq:fit_fc}
\end{equation}
which takes into account the build-up and decay of the $^{213}$Bi background activity.
The free parameters in the fit were the amplitudes of the counts from $^{221}$Fr, $A$, and from $^{213}$Bi, $D$, as well as  the half-life of $^{221}$Fr, $C$. The fixed parameter $B$ expresses the start of the ingrowth of the $^{213}$Bi activity and is determined by a fit of the ingrowth curve of the 440~keV $\gamma$ line in the $\beta$ decay of $^{213}$Bi. For the parameter $E$, the above mentioned literature value of the $^{213}$Bi half-life was used \cite{NNDC1991Martin}. A typical example of a fit to determine the half-life of $^{221}$Fr using Eq.~(\ref{eq:fit_fc}) is shown in Fig.~\ref{fig:221Fr_decayFit_corr}.
\\
For the integration of the peaks in the $\gamma$ spectrum, a linear background subtraction was carried out taking into account the amplitude of the background to the left and right of the peak. As the integral of the 218~keV $\gamma$ line in the decay of $^{221}$Fr went to zero, a single exponential could be fitted to extract the half-life (Fig.~\ref{fig:221Fr_GammadecayFit_corr}).
\\
All fits were performed using the MINUIT package \cite{James1975} inside the publicly available ROOT framework. Nearly all fits resulted in reduced $\chi^2$ values, $\chi^2/\nu$, close to unity. This also demonstrates that the integrations and dead-time corrections were being dealt with correctly. Whenever the $\chi^2/\nu$ of the fit was larger than unity the error on the corresponding $t_{1/2}$ was multiplied by $\sqrt{\chi^2/\nu}$.
\\
\begin{figure*}[!hbt]
\centering
\includegraphics[width = 0.75\textwidth]{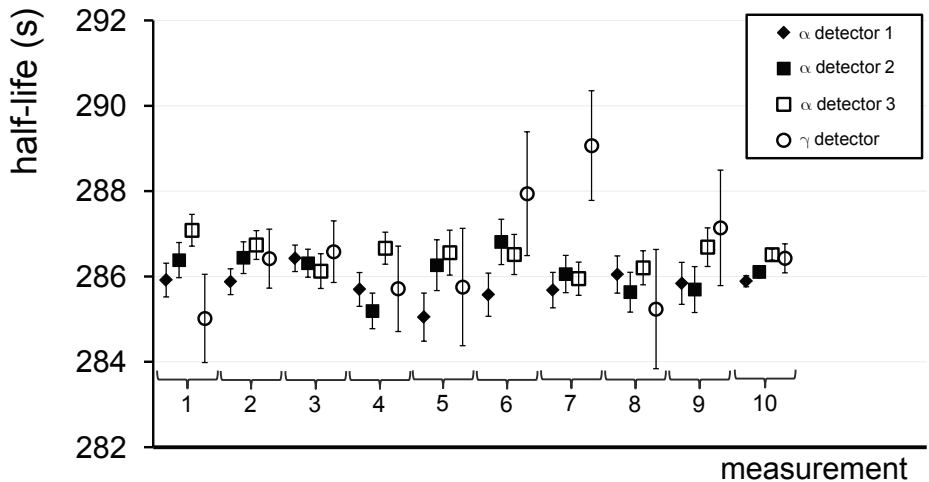}
\caption {$^{221}$Fr half-life values obtained in the 9 measurements at milliKelvin temperatures in the Au host foil, for the three $\alpha$~particle detectors (using the most intense $\alpha$~lines in the decay of $^{211}$Fr) and the 218~keV $\gamma$ line. The last block (labeled '10') shows the weighted average values for all 9 results for each detector.} \label{fig:221Fr_mKdata}
\end{figure*}
\begin{figure}[!hbt]
\centering
\includegraphics[width = 0.9\columnwidth]{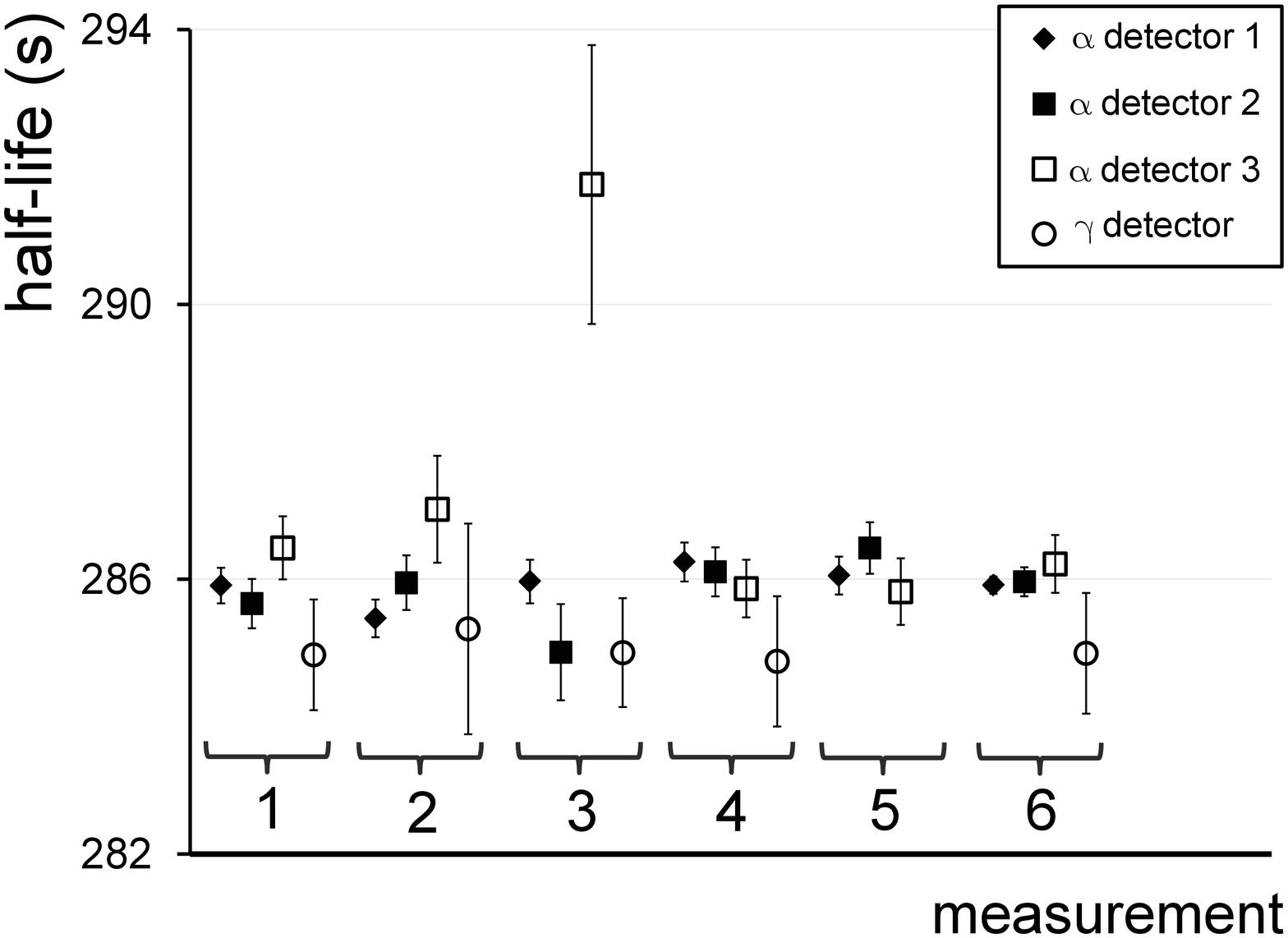}
\caption {$^{221}$Fr half-life values obtained in the 5 measurements at 4~K in the Au host foil. The last block (labeled '6') shows the weighted average values for all 5 results for each detector.} \label{fig:Plot4KAu}
\end{figure}
\begin{figure}[!hbt]
\centering
\includegraphics[width = 0.9\columnwidth]{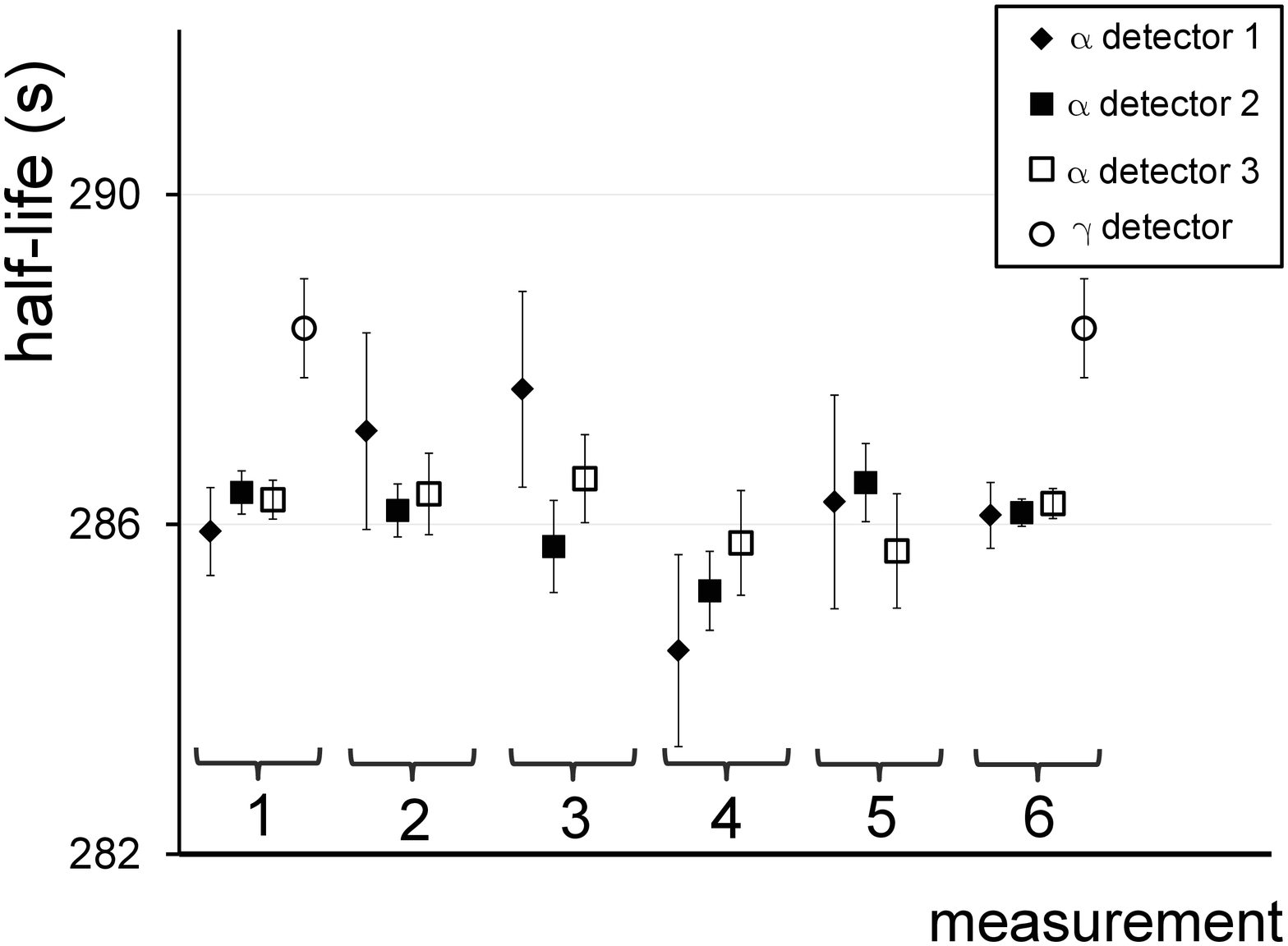}
\caption {$^{221}$Fr half-life values obtained in the 5 measurements in the Si host foil. During the last four measurements of this series, the gamma spectra were corrupted. The last block (labeled '6') shows the weighted average values for all 5 results for each detector.} \label{fig:Plot4KSi}
\end{figure}
The results from all four detectors, three $\alpha$~particle detectors and one $\gamma$~detector, as well as the results of successive decay measurements in one given material and for a given temperature, all turned out to be mutually consistent, thus demonstrating the stability of our experiment. This is illustrated in Figs.~\ref{fig:221Fr_mKdata}, \ref{fig:Plot4KAu}, and ~\ref{fig:Plot4KSi}, where all $t_{1/2}$ values extracted from the three different data sets are shown.
The evolution of the temperature of the sample during one of the measurements at milliKelvin temperatures is shown in Fig.~\ref{fig:Temperature_evolution}.
\begin{table*}[!hbtp]
\caption{\label{tab:Halflife_Fr}Half-life values for $^{221}$Fr measured by observing the decay $\alpha$~and $\gamma$~rays at different temperatures, i.e. in the mK~region or at 4~K, and in different environments, i.e. Au and Si. The values presented in the fifth column are the weighted averages of the values of all four detectors. The last column indicates the difference of the half-life values obtained, compared to the reference Si measurement. The quoted errors are purely statistical. Whenever the $\chi^2 / \nu$ of the weighted average was larger than unity, the error bar was increased by a factor $\sqrt{\chi^2 / \nu}$.}
\begin{ruledtabular}
\begin{tabular}{c|c|c|c|c|c|c}
    Measurement condition & Detector    &   t$_{1/2}$($^{221}$Fr) (s) & $\sqrt{\chi^2 / \nu}$&   t$_{1/2}$($^{221}$Fr) (s) & $\sqrt{\chi^2 / \nu}$ & $\Delta t_{1/2}$  (\%)\\
    \hline
     mK / Au    &   $\alpha_1$  &   285.89(13) &   0.8 &&& \\
                &   $\alpha_2$  &   286.10(11)  &   1.2 &&&\\
                &   $\alpha_3$  &   286.52(13)  &   0.8 &&&\\
                &   $\gamma$    &   286.43(36)  &   1.1 &&&\\
                &               &               &       &   286.17(8)   & 0.9 & 0.01(5)\\
    4~K / Au   &    $\alpha_1$  &   285.91(13)  &   1.2 &&&\\
               &    $\alpha_2$  &   285.96(18)  &   1.4 &&&\\
               &    $\alpha_3$  &   286.22(43)  &   3.0 &&&\\
               &    $\gamma$    &   284.92(95)  &   3.6 &&&\\
               &                &               &       &   285.93(11)  &   1.4 &   0.09(6)\\
    4~K / Si   &    $\alpha_1$  &   286.11(40)  &   1.2 &&&\\
               &    $\alpha_2$  &   286.14(21)  &   1.5 &&&\\
               &    $\alpha_3$  &   286.25(25)  &   0.9 &&&\\
               &    $\gamma$    &   288.38(60)  &   -   &&&\\
               &                &               &       &   286.19(12)  &   1.1 &   -\\
\end{tabular}
\end{ruledtabular}
\end{table*}
%
%
%
%
%
\\
Table~{\ref{tab:Halflife_Fr}} summarizes the results for the three different measurement conditions. Within 2.5 standard deviations, no difference is observed between the half-lives of $^{221}$Fr embedded in the non-metallic Si environment at 4~K, in Au at 4~K, and in Au at milliKelvin temperatures.
Our results therefore do not support the claimed temperature or host material dependence for $\alpha$~decay. The weighted average value of all three results in Table \ref{tab:Halflife_Fr} is $t_{1/2}$($^{221}$Fr) = 286.12(9) ($\chi^2/\nu = 1.6$).\\
%
%
The quoted error on this value is purely statistical. To quote an absolute half-life of $^{221}$Fr, systematic errors have still to be taken into account.
The systematic error resulting from the dead-time correction by pulser normalization was estimated by varying the markers which define the integral of the pulser peak, yielding a 0.5~s variation on the half-life of $^{221}$Fr. In addition, the integrals during the first half-life of $^{221}$Fr were the most sensitive to the dead-time correction, as at later times the amount of dead time rapidly dropped below 1~\%. Varying the starting point of the fit reveals another systematic error, related to the dead-time correction, of 0.7~s.
The systematic error related to the presence of counts from the decay of $^{213}$Bi was estimated to be 0.5~s by varying the value used for the half-life of $^{213}$Bi, i.e. parameter $B$ in Eq.~(\ref{eq:fit_fc}), within one standard deviation.
Finally, an 0.11~s systematic error comes from the accuracy of our timing system which was determined by calibrating it with a external clock.
\\
Adding all systematic errors in quadrature, our final value for the half-life of $^{221}$Fr becomes:
\begin{equation}
t_{1/2}(^{221}Fr)\; =\; 286.1\pm(0.09)_{stat}\pm(1.0)_{syst}\;s\;\;.
\label{eq:221Fr}
\end{equation}
This value is in agreement with and more precise than the value of 294(12)~s quoted in Refs. \cite{Audi2003,NNDC1991Martin}, and is in agreement with and even slightly more precise than the value of 287.4(12)~s quoted in Ref. \cite{Jeppesen2007}.

\section{Conclusion}

Because of the findings reported in \cite{Severijns2007} and \cite{Stone2007b}, it comes as no surprise that our results are not supporting the Debye-H\"{u}cker model \cite{Raiola2005,Kettner2006,Rolfs2006}, which predicts (based on the calculations of \cite{Ruprecht2008,Ruprecht2008b}) a reduction of the half-life of $^{221}$Fr in a metal at 4~K with a factor of about 50, and at 20~mK with a factor of many orders of magnitude more. No dependency on the solid-state environment and temperature of this $\alpha$~decaying isotope is observed up to a level of 1 $\times$ 10$^{-3}$, which is at variance with the reported 6~\% change of the activity of $^{213}$Po nuclei implanted in Cu at 12~K \cite{Raiola2007}. Our result supports the theoretical calculations of \cite{Zinner2007}, which predicted no significant change of the half-life. These dedicated half-life measurements bring down the level to which no temperature and host-material effect is observed on nuclear half-lives for $\alpha$~decays to the same level of precision as was already established for $\beta$ decays and EC decays in Refs. \cite{Nir-El2008,Goodwin2009,Kumar2008,Ruprecht2008,Ruprecht2008b,Fallin2008,Farkas2009}.

\section{Acknowledgments}

This work was supported by the Fund for Scientific Research
Flanders (FWO), project GOA/2004/03 of the K. U. Leuven, the Interuniversity Attraction Poles Programme, Belgian
State Belgian Science Policy (BriX network P6/23), and the grant LA08015
of the Ministry of Education of the Czech Republic. We thank Karsten Riisager, Hans Fynbo, Luis Fraile, and the ISOLDE collaboration for their support.




%

\end{document}